# VERY HIGH ENERGY GAMMA-RAYS FROM AGN: CASCADING ON THE COSMIC BACKGROUND RADIATION FIELDS AND THE FORMATION OF PAIR HALOS

F.A. Aharonian,[1] P.S. Coppi,[2] and H.J. Völk[1]

[1] Max-Planck-Institut für Kernphysik, D-6900 Heidelberg, Germany
[2] The Enrico Fermi Institute, University of Chicago, 5640 S. Ellis Ave, Chicago, IL 60637, USA

## ABSTRACT

Recent high energy $\gamma$-ray observations ($E_\gamma > 100$ MeV) of blazar AGN show emission spectra with no clear upper energy cutoff. AGN, considered to be possible sources for the observed flux of cosmic rays beyond $10^{19}$ eV, may well have emission extending into the VHE (very high energy, $E_\gamma > 100$ GeV) domain. Because VHE $\gamma$-rays are absorbed by pair production on the intergalactic background radiation fields, much of this emission may not be directly visible. The electromagnetic cascade initiated by the absorbed VHE $\gamma$-rays, however, may be observable. Since, most probably, the velocities of $(e^+, e^-)$ pairs produced in the cascade are quickly isotropized by an ambient random magnetic field, extended "halos" ($R > 1$ Mpc) of pairs will be formed around AGN with VHE emission. The cascade radiation from these pair halos is emitted isotropically and should be observable at energies below a few TeV. The halo radiation can be distinguished by its characteristic variation in spectrum and intensity with angular distance from the central source. This variation depends weakly on the details of the central source model, e.g., the orientation and beaming/opening angle of an emitting jet. Limiting or determining the intensity of the pair halo can thus serve as a model-independent bound on or measure of the VHE power of AGN. Next-generation Cherenkov telescopes may be able to image a pair halo.

*Subject Headings:*  radiation mechanisms: radiative transfer — cosmology:diffuse radiation — gamma-rays: theory — galaxies: active

## I. INTRODUCTION

To date, the EGRET instrument on the Compton Gamma-Ray Observatory (GRO) has detected over 20 AGN at energies up to $\sim 10$ GeV (Fichtel *et al.* 1993). The very high apparent $\gamma$–ray luminosities of these objects, their observed variability on timescales as short as a few days, and the fact that all EGRET AGN appear to be blazars thought to have a relativistic jet pointing in our direction, strongly suggest that their $\gamma$-ray emission is beamed and originates in the relativistic jet (Kniffen *et al.* 1993). Whether or not the spectra of these AGN continue unbroken beyond 10 GeV is a key question for models of the $\gamma$-ray emission and AGN jets. (Unfortunately, because of the typically low photon statistics at 10 GeV, EGRET cannot address this issue.) AGN are recognized as a potential class of cosmic ray sources with spectra extending up to $E \geq 10^{20}$ eV (e.g., see Kafatos *et al.* 1981), and they contain enough target material for the efficient production of $\gamma$-rays via Inverse Compton (IC) or proton-proton and proton-photon interactions. It thus seems natural to expect $\gamma$-ray emission beyond the energy domain of EGRET. At least in one case, Mrk 421, the spectrum appears to extend up to 1 TeV (Punch *et al.* 1992), though it is not obvious that this radiation is directly connected with the low energy part of $\gamma$-radiation detected by EGRET. For VHE $\gamma$-rays ($E_\gamma \geq 100$ GeV), the problem of absorption inside the source becomes severe. For example, the roughly isotropic (unbeamed) emission of an AGN accretion disk at UV and optical frequencies (the so-called "blue bump") implies a large absorption probability due to photon-photon pair production for VHE $\gamma$-rays emitted near the accretion disk (Coppi *et al.* 1993). Thus, we might hope to see VHE $\gamma$−rays from AGN provided their production takes place far away from the central compact source. (This concern may not apply to relatively weak sources like Mkn 421 which never show a "blue bump" in their spectra). This requires an efficient transport of nonthermal energy, e.g., by cosmic rays, from the central engine to large distances ($\gtrsim$ 1 pc) if we believe that the main energy release takes place near the accreting black hole.

Absorption of VHE $\gamma$-rays by photon-photon pair production with low energy photons is important not only inside sources. Intergalactic radiation fields such as the 2.7 K microwave background radiation (MBR) represent a non-negligible absorption opacity when cosmological propagation distances are considered (Gould and Schréder, 1966; Jelley, 1966). The MBR, for example, is not transparent to extragalactic $\gamma$-rays with energy $E_c \gtrsim$ 100 TeV. The spectral cutoff energy depends on the distance to the source and the character of the evolution of the MBR; hence it contains unique cosmological information (Aharonian & Atoyan 1985). As noted in Stecker *et al.* (1992), pair production of lower energy $\gamma$-rays on the background far IR and optical ($IR/O$) fields may also be extremely important and could reduce $E_c$ for distant sources down to 100 GeV. The value of $E_c$ depends strongly on the $IR/O$ background level, and therefore is very uncertain (see Fig. 1). When a VHE $\gamma$−ray is absorbed, however, its energy is not lost. The pairs produced in the interaction will create new $\gamma$-rays by inverse Compton scattering on the background field photons. These secondary $\gamma$-rays can in turn be absorbed to create more pairs, and an electrogmagnetic cascade develops. The cascade ends when the Compton upscattered photons no longer have enough energy to pair produce on the background photons. In effect, the cascade has transformed the energy of the initial photon into many photons of much lower energy. Unlike the original VHE $\gamma$−ray, these low energy photons will be observable. In this letter, we study the radiation from cascades initiated by VHE $\gamma$-rays and argue that it is an important, potentially detectable, signature of VHE emission from AGN.

## 2. CASCADE RADIATION FROM PAIR HALOS

The idea that cascade processes initiated by $\gamma$- or cosmic rays play an important role in the intergalactic medium (IGM) is not a new one (e.g., Wdowczyk *et al.* 1972, Aharonian & Atoyan 1985, Protheroe 1986, Zdziarski 1988, Protheroe & Stanev 1993). All previous calculations, however, assume that the cascade particles travel in a straight line to the observer. For calculations of the diffuse background resulting from isotropic VHE sources uniformly distributed in space, this is not a bad first approximation. One must be more careful, though, when dealing with the radiation from a single source. Not only must one consider the physics of the production of secondary particles in the cascade, one must also consider how secondary pairs propagate in the intergalactic magnetic field. Unfortunately, our knowledge of the strength of the intergalactic magnetic field and its fluctuations on different scales is limited. A crude estimate obtained by equating the magnetic pressure to an average intergalactic pressure of $6.75 \times 10^{-16}$dyne/cm$^2$ (Breitschwerdt et al., 1991) suggests the field could be as high $B \sim 10^{-7}$ G. Even for a much smaller field of $10^{-9}$G, the corresponding gyroradius for pairs of energy $E$ is only $\sim 100(E/100\text{TeV})(B/10^{-9}\text{ G})$ pc. For pairs of energy $E \lesssim$ 100 TeV, this is much less than the Compton cooling length, $\Lambda_c = E/(dE/dx) \sim 4 \times 10^3 (E/100 \text{ TeV})^{-1}$ pc. As a consequence, those pairs with pitch angle $\alpha \gtrsim \pi/4$ will Compton upscatter photons more or less isotropically. While the exceptional $\alpha \sim 0$ pairs will not do so, these pairs diffuse in pitch angle (and thus in spatial directions) due to turbulent fluctuations in the field. Taking the field to be ordered on scales less than the mean distance between galaxies, $\sim$ 2 Mpc, we estimate a diffusion coefficient $D \sim 4 \times 10^{32}(E/100 \text{ TeV})^{0.5}(B/10^{-9}\text{ G})^{-0.5}$cm$^2$/s and a mean free path $\lambda_e = 3D/c \sim 10(E/100\,\text{TeV})^{0.5}(B/10^{-9}\,\text{G})^{-0.5}$ kpc. Even $\alpha \sim 0$ pairs with $E \lesssim 100\,\text{TeV}$ are thus expected to undergo large angle deflections on scales comparable to or much less than their Compton cooling lengths. Note also that while pairs are cooling, they diffuse a negligible distance ($\sim \min[(\Lambda_c\lambda_e)^{1/2}, \Lambda_C]$) compared to the distance to a cosmological source ($\gtrsim 100$ Mpc). At least for pairs of energy $\lesssim 100$ TeV, it then seems reasonable to make the following simplifying assumptions: (i) cascade pairs do not propagate from the site of their creation, (ii) their velocities are instantly isotropized, and (iii), they consequently Compton upscatter photons isotropically into all directions. Although assumptions (ii) and (iii) may not hold at energies $\gtrsim 100$ TeV, photons upscattered by these pairs have energies $\gtrsim 25$ TeV and are typically not observable because of the strong intergalactic absorption. In practice, then, it is usually a good approximation to treat pairs with $E \gtrsim 100$ TeV in the same way as pairs with $E \lesssim 100$ TeV.

In this case, one sees that the cascading started by a VHE source effectively surrounds the source with a giant, isotropically emitting *halo* of pairs. Because of the very rapid decrease in the photon absorption length $\lambda_{\gamma\gamma}(E)$ for photon energy $E \gtrsim 100$ TeV (see Fig. 1c), the cascade around the source can be considered to occur in two stages. In the first, primary photons with $E > 100$ TeV pair produce on microwave background photons and start a Klein-Nishina cascade. Because of the short interaction lengths at these energies, the cascade on the microwave photons ends within a very short distance of the source, on a length scale $L_1$ typically $\lesssim 1$ Mpc for $E < 10^{17}$ eV. The emission produced in this stage has an intensity simply proportional to the source luminosity above $\sim 100$ TeV and a standard spectrum independent of the primary source spectrum (see Aharonian & Atoyan 1985, Protheroe 1986, and Zdziarski 1988). In the second stage, the cascade photons produced in the first stage, along with source photons of energy $E \lesssim 100$ TeV, initiate a Thomson cascade that is mediated by *both* the microwave and $IR/O$ background photon fields. Because of the energy threshold for pair production, photons of energy less than 100 TeV can pair produce only on $IR/O$ photons. The Compton scattering of secondary pairs, however, is still dominated by the MBR which has a much higher energy and number density than the $IR/O$ background radiation. In this second stage, then, $\gamma$-rays of energy $E_0$ produce secondary pairs of mean energy $\sim E_0/2$. These secondary pairs in turn Compton upscatter photons to typical energies $E_\gamma \sim \frac{4}{3}(E_0/2m_ec^2)^2(2.7kT_{MBR}) \sim 1(E_0/40\,\text{TeV})^2$ TeV ($\ll E_0$). With this information, one can predict how the inverse Compton emission from the pair halo will appear to a distant observer. If the VHE source spectrum contains a significant amount of luminosity above 100 TeV, then at all energies below the $IR/O$ absorption cutoff, an observer a distance $d$ from the source will see a narrow core of emission around the source that has angular size $\sim L_1/d$ ($\lesssim .5°$ for $d \gtrsim 100$ Mpc). This emission comes from the energetic pairs produced in the first stage of the cascade. As the observer enlarges the opening angle of his detector, he will begin to see radiation from pairs produced in the second stage of the cascade. Since $\lambda_{\gamma\gamma}$ decreases rapidly with increasing energy, the pairs responsible for observed photons of energy $E_\gamma$ must have been created within a distance $L_2 \sim \lambda_{\gamma\gamma}(E_0)$ of the source. Hence, at energies $E > E_\gamma$, the pair halo produced in the second stage appears as an extended emission region of angular size $\sim L_2/d$.

To estimate this angular size or calculate the actual halo emission spectrum observed, we must make some assumptions about the shape and level of the $IR/O$ background spectrum. While the MBR is thought to be well-understood, current knowledge of the extragalactic $IR/O$ background fields is mainly theoretical. Consequently, the free paths of $\gamma$-rays in these fields are rather uncertain. We have chosen to follow Stecker *et al.* (1992) and use two representative $IR/O$ backgrounds in our calculations: a "high" intensity one, $n(\epsilon) = 1.5 \times 10^{-3}\epsilon_{eV}^{-2.55}\,\text{cm}^{-3}\text{eV}^{-1}$ for $\epsilon > 0.02\,\text{eV}$, which the authors suggest as an upper limit to the $IR/O$ background intensity, and a "low" intensity one, $n(\epsilon) = 8 \times 10^{-4}\epsilon^{-2}\,\text{cm}^{-3}\text{eV}^{-1}$ for $\epsilon > 0.02$eV which the authors suggest as a lower limit. (The recent analysis by Stecker & De-Jager 1993 of the Whipple Mrk 421 data, in fact, gives an upper limit that is slightly lower than the "high" level used here.) The resulting photon absorption lengths and the corresponding absorption spectra are shown in Fig. 1. For observed energies $E > 1$ TeV and source distance $d = 1000$ Mpc, these free paths give (second stage) halo emission regions of apparent angular size $\sim 3$ and $0.3$ degrees respectively for the "low" and "high" $IR/O$ background levels.

To calculate examples of what the halo emission spectrum might look like, we have used a standard Monte Carlo code. The code follows a photon until it pair produces and records the position where the pairs are produced. The code then follows pairs in energy as they cool and Compton upscatter photons. The upscattered photons are then followed to determine where they pair produce, etc. The process continues until all surviving photons have mean free paths greater than some suitably large distance (10,000 Mpc). The pair distribution, and thus the halo emissivity, as a function of distance from the source is then obtained by binning up the pair positions into logarithmically spaced radial bins with a resolution of 10 bins per decade of radius. (Photons were assumed to be emitted isotropically from the source so we could the exploit the spherical symmetry of the problem.) The flux measured by an observer at an arbitrary position is calculated by shooting rays through the halo and explicitly evaluating the angular integral $F_\nu = \int J_\nu(\mu) d\mu$. For the calculations shown, we used the exact interaction cross-sections, but we chose to ignore cosmological corrections (e.g., the change in background photon energy with redshift) since they should not change qualitatively the results presented, and the level of the $IR/O$ background as a function of redshift, a critical quantity, depends on galaxy evolution scenarios and is not well-known. Sample halo spectra, calculated for different source distances, background levels and primary energies, are shown in Fig. 2.

## 3. DISCUSSION

As can be seen from Fig. 2, the spectrum of radiation observed to come from a pair halo has a roughly standard shape: $J_\gamma(E) \propto E^{-1.5}$ at $E \ll E^*$, with steepening near $E \sim E^*$, and a cutoff beyond several times $E^*$. The value of $E^*$ depends on the source distance: $E^* \geq 1$ TeV for $d = 100$ Mpc, and $E^* \leq 100$ GeV beyond 1000 Mpc. In general, the lower the level of $IR/O$ background, the harder the spectrum and the larger the angular size of the halo emission. Also, the lower the energy of a primary, the longer its absorption length, and the more diffuse the halo emission it contributes to appears. The spectral and angular dependence of the observed halo radiation is thus, in principle, a function of the VHE primary source spectrum. We note, though, that if the VHE source luminosity is dominated by primaries with $E \gtrsim 100$ TeV, the halo radiation produced in the second stage of the cacade is dominated by reprocessed photons from the first stage of the cascade. Since first stage photons are produced very near the source and have a spectrum independent of the primary spectrum, the input to the second stage cascade is then insensitive to the primary source spectrum, and consequently, so is the spectrum observed from the halo. As one integrates over a larger patch of sky, the halo spectrum eventually approaches and, at low energies, exceeds the spectrum expected from a "direct" (rectilinear propagation) cascade. This may be understood by noting that since $\lambda_{\gamma\gamma}$ can be extremely large, the observer is actually sitting *inside* the pair halo. At low energies $E_\gamma$, for which $L_2(E\gamma)$ exceeds the source distance $d$, the observer sees cascade photons coming from *all* directions in the sky, not just from the source. This is a potentially important correction to low energy diffuse background calculations based on rectilinear cascade propagation since the observer detects many more photons from nearby VHE sources.

To assess the importance of the halo emission relative to the direct unabsorbed source emission, $I(E) = I_0(E) \exp(-d/\lambda_{\gamma\gamma})$, we show in Fig. 1a,b the total spectra observed within 5° of the source (computed including the effects of cascading) for a primary source spectrum $I_0(E) \propto E^{-2}$ extending up to 500 TeV. For such an $I_0(E)$, the curves $\exp(-d/\lambda_{\gamma\gamma})$ correspond exactly to the primary source "absorption" spectrum, $E^2 I(E)$. In this example, the halo contribution appears only as an excess at the highest energies. However, for harder primary spectra or primary spectra with a high energy cutoff $\gg 500$ TeV, the halo emission could dominate entirely over the direct source emission. Of course, if one can look off the source or the direct radiation is beamed away from the observer, one will see only the halo emission. Note also that unlike the emission from the central point source, the emission from a halo should not be time variable because of the large distance scales involved. The possibilities for detecting a halo could thus be enhanced by waiting for the central source to drop into a low state. (The halo emission represents a minimum flux level below which the overall source emission should not drop.)

The total flux of $\gamma$-radiation from the halo depends mainly on the total luminosity and duty cycle of the source at energies $\gtrsim 10$ TeV (the lower energy primaries contribute only to the outermost part of the pair halo which is difficult to detect), but not on the geometry of the source, in particular the direction and beaming/opening angle of the jet emitting the $\gamma$-rays. This creates a unique possibility of verifying the jet origin of $\gamma$-rays from AGN. While the 100 MeV emission may be strongly beamed, the halo emission is not. Many more AGN may be visible in the VHE range, and if one can extrapolate or relate the observed VHE power to the true source power at lower energies, one may obtain a direct measure of the jet beaming factor. To carry out this procedure, one must determine the total energy flux of the source integrated over the typical angular size of the halo. This depends on the distance to the source and the level of the $IR/O$ background of the local enviroment. Thus, the investigation of the angular and spectral distribution of halo radiation could, in principle, provide source model-independent and almost unambiguous information about the universal $IR/O$ background, the source distance, and the VHE power of AGN.

The minimum energy flux of photons above $100\,\text{GeV}$ detectable by modern imaging Air Cherenkov Telescopes (ACT) could be as low as $\approx 1\,\text{eV}/\text{cm}^2\text{s}$ for relatively compact sources of angular size $\leq \Delta\psi \approx 10$ arcminutes. VHE $\gamma$-rays could then be detected from the central region of the pair halo, e.g., within $\leq 0.2°$ of the source, provided the AGN source power was $L_0(\geq 10\,\text{TeV}) \gtrsim 3 \cdot 10^{45}(d/10^3\,\text{Mpc})^2)\,\text{erg/s}$ (see Fig. 2). For 3C 279 this represents a fraction $\approx 10^{-2} f^{-1}$ of the source's observed 100 MeV-10 GeV luminosity ($\sim f \times 10^{48}\,\text{ergs/s}$), where the beaming factor $f$ is thought to be within $10^{-2}$ and $10^{-3}$. So far VHE $\gamma$-radiation has been detected from only one extragalactic source, namely Mrk 421 (Punch et al., 1992). The flux of Mrk 421 above 500 GeV is roughly $10\,\text{eV}/\text{cm}^2\text{s}$. From a comparison of this value to the spectra in Fig. 2, we conclude that the Whipple flux may, in principle, be explained by halo radiation provided the primary VHE luminosity of Mrk 421 exceeds $10^{44} - 10^{46}$ erg/s. The measured flux, of course, could simply be the "direct" radiation from the jet. (An observation of time variability would limit the possible contribution from a halo.) Note that while the hard spectrum of halo $\gamma$-rays extends down to lower energies, the flux expected at 100 MeV is well below the EGRET detection of $J_\gamma(E > 100\,\text{MeV}) = (1.4 \pm 0.3) \times 10^{-7}\,\text{ph/cm}^2\text{s}$ (Lin et al, 1993), i.e., the high energy radiation observed by GRO is almost certainly not connected with a pair halo.

For a firm detection of the halo radiation, information on the time as well as spectral and angular distributions of the radiation is required. Obviously, the most convincing evidence for the formation of pair halos around AGN would be the detection of the VHE $\gamma$ emission from a region of angular size $\gtrsim 1°$ by future wide-angle ACT.

PSC gratefully acknowledges support from a GRO Fellowship.

## Figure Captions

**Figure 1.** The top two panels (Fig. 1a,b) show the changes in the observed $\gamma$-ray spectra ($I_\gamma/I_0$) due to absorption by pair production on the MBR and IR/O background fields for several source distances, $d$. The topmost and middle panels are calculated assuming respectively the "high" and "low" level of IR/O background discussed in the text. For comparison, the heavy solid lines show the spectra ($I_{cas}/I_0$) obtained when the effects of cascading (see text) on the background radiation are included. (In this case, the primary VHE photons had a power-law spectrum, $I_0 \propto E^{-2}$, extending in energy up to $5 \times 10^{14}$ TeV, and the source was at a distance of 120 Mpc. The halo radiation emitted within 5° of the source is shown.) The bottom panel (Fig. 1c) shows the mean free paths, $\lambda_{\gamma\gamma}$, for $\gamma$−rays in various background radiation fields in the IGM. The dashed horizontal lines are plotted at the distances of some possible VHE sources.

**Figure 2.** The cascade $\gamma$-ray fluxes, measured within various ranges of angles from the central VHE source, that are expected from the extended pair halos discussed in the text. The *solid*, *dotted*, and *dashed* curves show respectively the flux received within 0.2°, 1°, and 3° of the source. For comparison, the *heavy solid* curves give the the "direct" cascade flux expected if the source were to emit isotropically and the cascade particles propagated in a straight line from the source (i.e., if the cascade pairs were not deflected by a magnetic field). The *heavy dotted* curves give the halo spectrum integrated over the entire sky. Fig. 2a shows the spectra obtained for monoenergetic primary source emission at $E_0 = 500$ TeV. The upper panel was calculated using the "high" level of $IR/O$ background, while the lower panel was calculated using the "low" $IR/O$ background level (see text). Fig. 2b shows the same case, except that the source emission is now at $E_0 = 50$ TeV. For both Fig. 2a,b, the VHE source luminosity was taken to be $L_0 = 10^{44}$erg/s, and the source was placed at a distance of 100 Mpc. Fig. 2c shows the spectra obtained for a "low" IR/O background level and the source at a distance of 1000 Mpc with VHE luminosity $L_0 = 10^{46}$erg/s; the upper panel shows the case of source emission at $E_0 = 500$ TeV, while the lower panel shows the case of source emission at $E_0 = 50$ TeV.

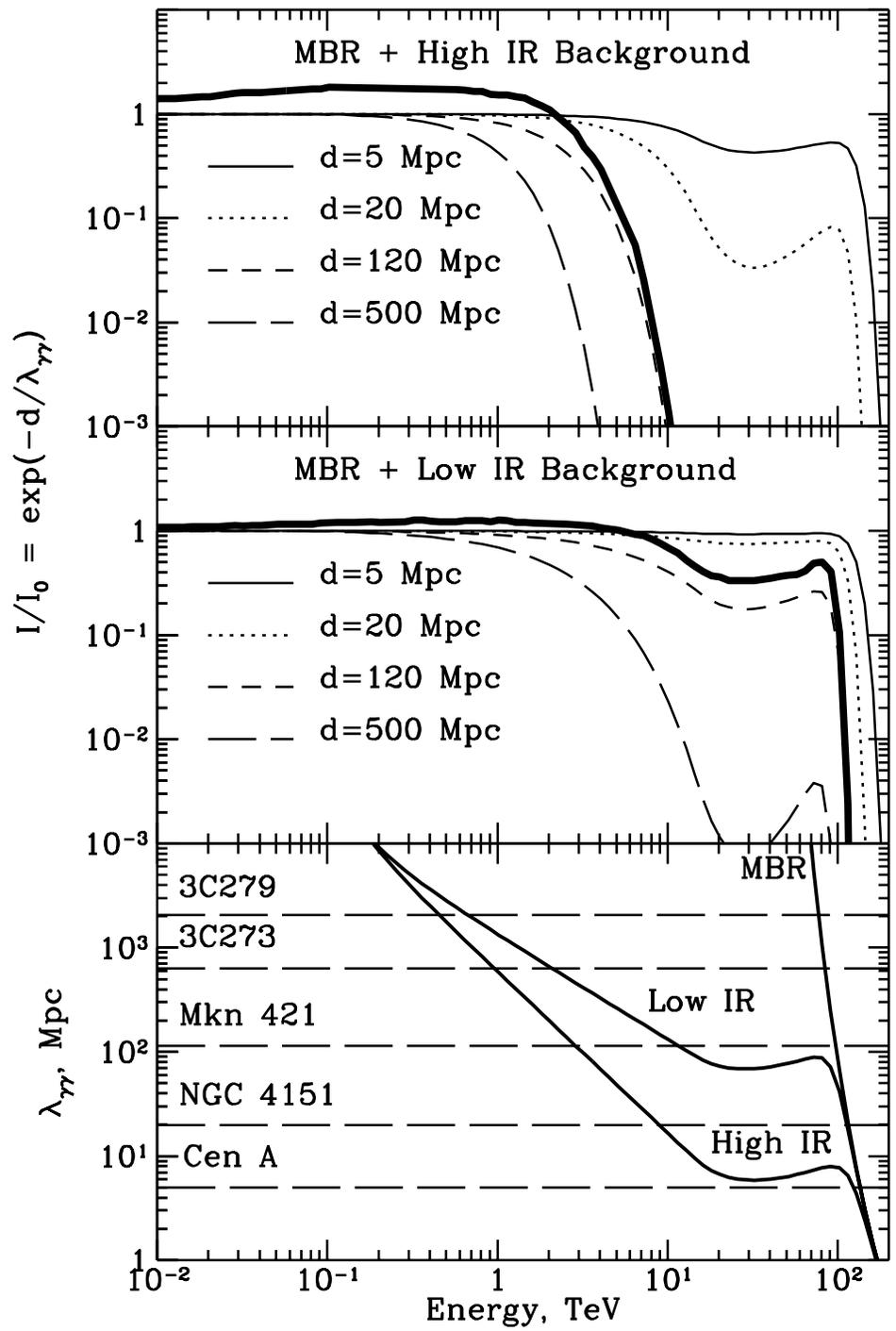

Fig. 1

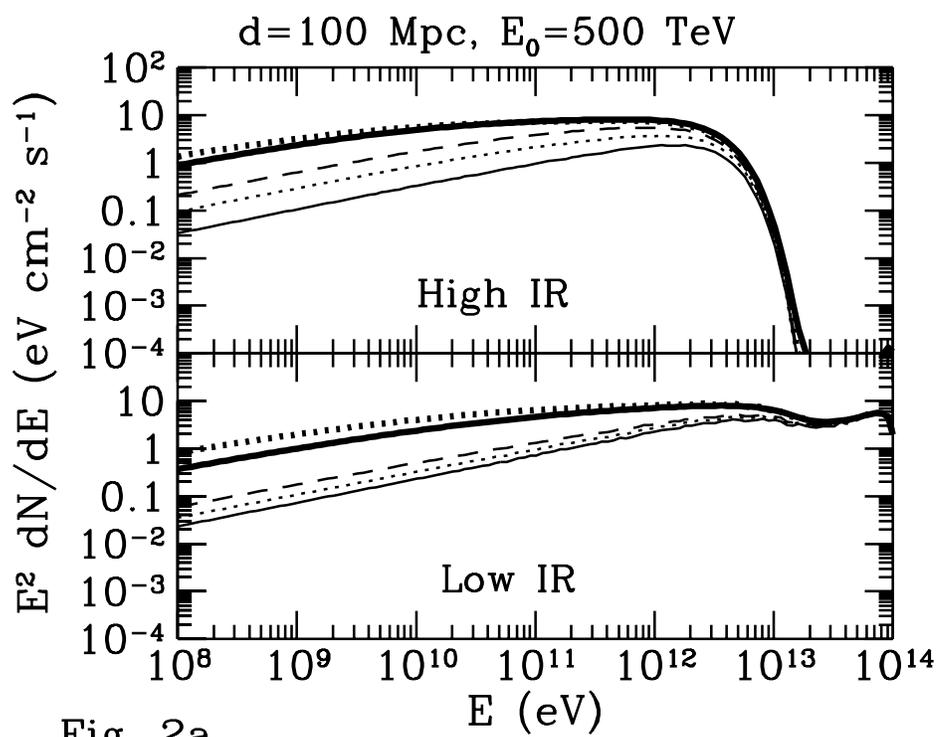
Fig. 2a

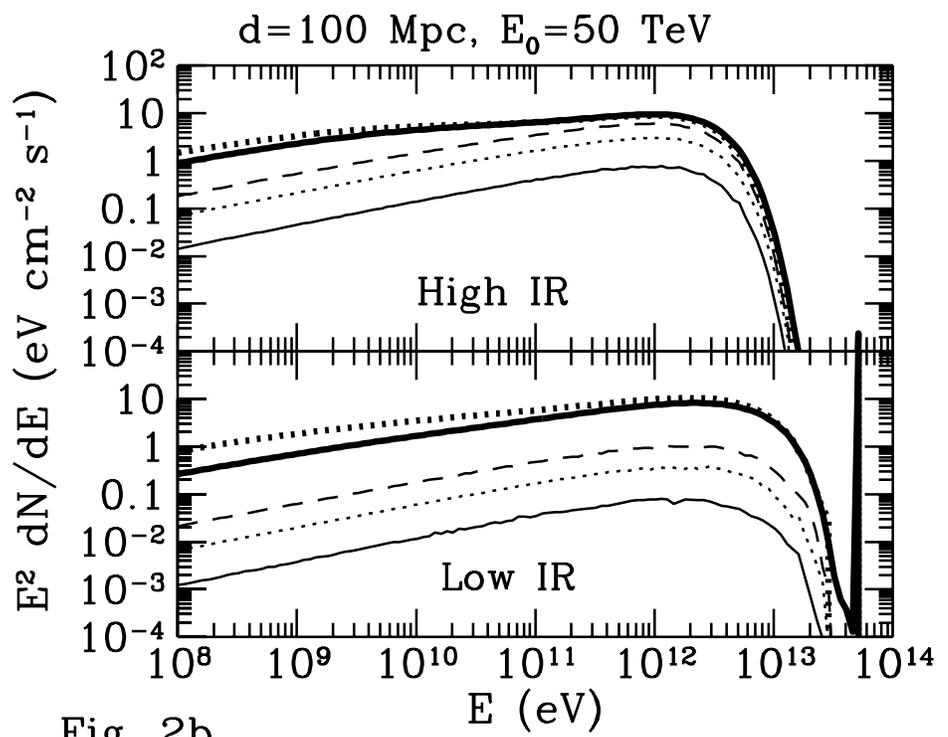

Fig. 2b

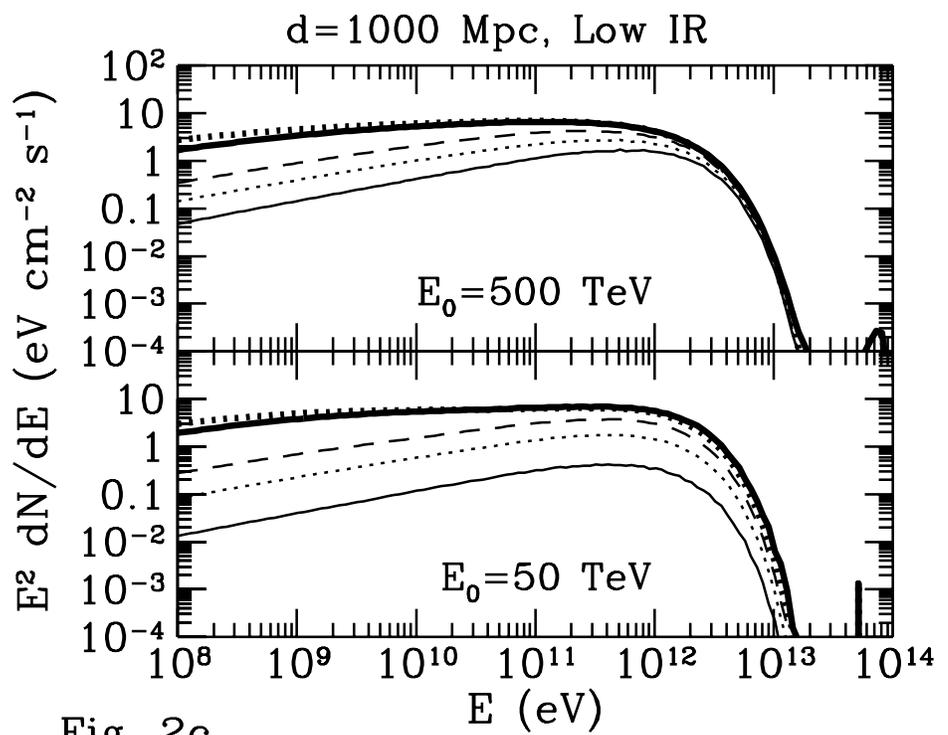

Fig. 2c